\documentstyle[twocolumn,aps,prl,epsf]{revtex}

\begin{document}
\draft

\twocolumn[
\hsize\textwidth\columnwidth\hsize\csname@twocolumnfalse\endcsname

\title{\bf Common trends in the critical behavior of the
Ising and directed walk models}
\author{Ferenc Igl\'oi\cite{permanent} and Lo\"{\i}c Turban}
\address{Laboratoire de Physique du Solide, Universit\'e
Henri Poincar\'e (Nancy I), B.P. 239,\\
F-54506 Vand\oe uvre l\`es Nancy Cedex,
France}
\date{\today}
\maketitle
\begin{abstract}
We consider layered two-dimensional Ising and directed walk models and show that
the two problems are inherently related. The information about the zero-field
thermodynamical properties  of the Ising model is contained into the transfer
matrix of the directed walk. For several hierarchical and aperiodic
distributions of the couplings, critical exponents for the two problems are
obtained exactly through renormalization.
\smallskip\\
{\sf cond-mat/9606118}   
\end{abstract}

%\pacs{05.50.+q, 64.60.Ak, 68.35.Rh}
\vglue.6truecm]

The Ising model (IM) and the directed walk (DW) are among the most studied
problems in lattice statistics. The IM is a standard model for magnetic or
liquid-gas phase transitions whereas the DW can be used to describe linear
fluctuating objects such as directed polymers, flux lines or
interfaces in two-dimensional systems. 

The IM is exactly solvable in two dimensions~\cite{onsager44} and the 
solution can be generalized for layered systems with different types of
distributions for the interlayer couplings such as periodic~\cite{periodic},
quasi-periodic~\cite{quasiperiodic}, aperiodic~\cite{luck93,aperiodic} and
random~\cite{random}. The DW is probably the simplest non-trivial problem in
statistical mechanics for which exact results can be obtained on
homogeneous~\cite{fisher86}, inhomogeneous~\cite{inhdw} and 
random~\cite{randdw} lattices.

In this Letter, we present a hitherto unnoticed connection between the IM and
the DW in two dimensions. Both problems are considered on layered lattices,
such that the walk is directed along the translationally invariant direction.
We show that the complete solution of the DW, i.e. the diagonalization of its
transfer matrix (TM), provides all the necessary information to obtain the
zero-field thermodynamical properties and correlation functions of the IM.  
The DW approach, which is  simpler, is then used to perform an exact
renormalization-group (RG) study of the TM eigenvalue problem for
self-similar distributions of the couplings. {\it The critical properties of
the IM and DW are governed by two different fixed points of the same
RG-transformation}. 

Let us first present the hidden relation between the two problems. We consider
a layered IM in the extreme anisotropic limit~\cite{kogut79}. The transfer
matrix going in the direction parallel to the layers is $\exp(-\tau{\cal H})$,
where $\tau$ is the lattice spacing in the Euclidian time direction, and ${\cal
H}$ the Hamiltonian of a quantum Ising chain: 
\begin{equation}
{\cal H}=-{1\over2}\sum_{k=1}^{L} h_k\,\sigma_k^z 
-{1\over2}\sum_{k=1}^{L-1} J_k\,\sigma_k^x\sigma_{k+1}^x\; . 
\label{e1}
\end{equation}
The $\sigma_i$s are Pauli spin operators, the transverse field $h_k=h$ plays
the role of the temperature and the couplings $J_k$ are non-periodic.

Following Lieb {\it et al}~\cite{lieb61} and Pfeuty~\cite{pfeuty70} ${\cal H}$
can be transformed into a free-fermion model 
\begin{equation}
{\cal H}=\sum_{q=1}^L\Lambda_q(\eta_q^\dagger\eta_q -{1\over2}) 
\label{e2}
\end{equation}
in terms of the fermion creation and annihilation
operators $\eta_q^\dagger$, $\eta_q$. The fermion excitations
$\Lambda_q$ are non-negative and satisfy the set of equations
\begin{eqnarray}
\Lambda_q\Psi_q(k)&=&-h_k\Phi_q(k)-J_k\Phi_q(k+1)\nonumber\\
\Lambda_q\Phi_q(k)&=&-J_{k-1}\Psi_q(k-1)-h_k\Psi_q(k)
\label{e3}
\end{eqnarray}
with the boundary conditions $J_0=J_L=0$. The ${\bf\Phi}_q$s and 
${\bf\Psi}_q$s, which are related to the coefficients of a canonical
transformation, are normalized. They enter into the expressions of correlation
functions and thermodynamical quantities~\cite{lieb61,pfeuty70}.

Usually one proceeds by eliminating either ${\bf\Psi}_q$ or ${\bf\Phi}_q$
in~(\ref{e3}) and the excitations are deduced from the solution of an  
eigenvalue problem. This last step can be avoided by introducing a
$2L$-dimensional vector ${\bf{V}}_q$ with components $V_q(2k-1)=-\Phi_q(k)$,
$V_q(2k)=\Psi_q(k)$ and noticing that the relations in Eq.~(\ref{e3})
correspond to the eigenvalue problem for the matrix 
\begin{equation}
T=
\left(\begin{array}{ccccccc}
0&h_1&0&0&0&0&\cdots\\
h_1&0&J_1&0&0&0&\cdots\\
0&J_1&0&h_2&0&0&\cdots\\
0&0&h_2&0&J_2&0&\cdots\\
\vdots&\vdots&&\ddots&&\ddots&
\end{array}\right)
\label{e5}
\end{equation}
which can be interpreted as the TM of a DW problem on two interpenetrating,
diagonally layered square lattices. The walker makes steps with
weights $h_k$ and $J_k$ between first-neighbour sites on one of the two
lattices.

Changing ${\bf\Phi}_q$ into $-{\bf\Phi}_q$ in ${\bf{V}}_q$, the eigenvector
corresponding to $-\Lambda_q$ is obtained. Thus all the information about the
DW and the IM is contained into that part of the spectrum with $\Lambda_q\geq0$.
Later on we shall restrict ourselves to this sector.

Let us now consider the correlation lengths in the direction parallel to the
layers for both problems. For the DW it can be expressed as a function of the
two leading eigenvalues of the TM with:
\begin{equation}
\xi_\parallel^{\rm DW}=\left[\ln\left({\Lambda_L\over\Lambda_{L-1}}
\right)\right]^{-1} \simeq{\Lambda_L\over\Lambda_L-\Lambda_{L-1}}\; .
\label{e6}
\end{equation}
Thus $\xi_\parallel^{\rm DW}$ is proportional to the inverse gap at the top of
the spectrum. For the IM in the disordered phase the correlation length is the
inverse of the lowest excitation energy of ${\cal H}$ in Eq.~(\ref{e2}) so that
\begin{equation} \xi_\parallel^{\rm IM}\sim\Lambda_1^{-1}\; .
\label{e7}
\end{equation}
$\Lambda_1$ is also the lowest eigenvalue in the spectrum of the TM. In the
ordered phase $\Lambda_1=0$ and the correlation length involves the second
eigenvalue $\Lambda_2$.

\begin{figure}
\epsfxsize=8.6cm
\begin{center}
\hspace*{-13truemm}\mbox{\epsfbox{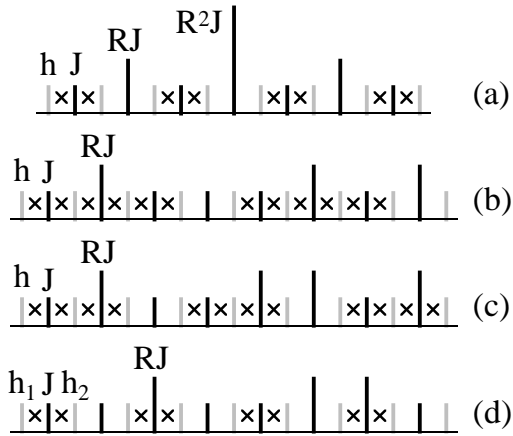}}
\end{center}
\caption{Matrix-elements $T_{j,j+1}$ as a function of $j$ for different 
non-periodic sequences: a) hierarchical b) period-doubling c) three-folding d)
paper-folding. Components of the eigenvector to be decimated out in the RG
transformation are denoted by crosses. The heights of solid vertical bars
indicate the strength of the couplings, the grey  bars stand for the field.} 
\end{figure}
 
Approaching one of the two critical points, the correlation length of the
problem is diverging and the corresponding part in the  TM spectrum displays a
scaling behavior. Let us consider a finite system with transverse size $L\gg1$
and denote by $\Delta\Lambda_i$ either $\Lambda_L-\Lambda_{L-i}$ for the DW or
$\Lambda_i$ itself for the IM with $i\ll L$. When lengths are rescaled by a
factor of $b>1$, i.e. with $L'=L/b$, the gaps are assumed to behave as
\begin{equation}  
(\Delta\Lambda_i)'=b^{y_{\Lambda}}\Delta \Lambda_i\; ,
\label{e8}
\end{equation}
where $y_{\Lambda}$ is the gap exponent which is generally different at
different parts of the spectrum. This leads to the finite size behavior
\begin{equation}
\Delta\Lambda_i(L)\sim L^{-y_{\Lambda}}\; ,
\label{e9}
\end{equation}
thus from Eqs.~(\ref{e6}) and~(\ref{e7}) the longitudinal correlation lengths
are $\xi_{\parallel}\sim L^{y_{\Lambda}}$. Since $\xi_{\perp}\sim L$, the
anisotropy exponent $z$, such as $\xi_{\parallel}\sim\xi_{\perp}^z$, is given
by $z=y_{\Lambda}$. For the DW one is interested in the transverse fluctuations
which are characterised by the wandering exponent $w$ through $\xi_{\perp}\sim
\xi_{\parallel}^w$, thus $w=y_{\Lambda}^{-1}$.

The scaling properties of the spectrum of $T$ are conveniently studied
using RG techniques. We consider different self-similar lattices for
which exact RG transformations can be worked out so that we obtain exact 
results about the critical properties of both the IM and the DW. In the
following the transverse field is assumed to be constant and equal to $h$.

\paragraph*{Hierarchical sequence}
We start with a hierarchical lattice in which the couplings $J_k$ follow the
Huberman-Kerszberg sequence~\cite{huberman85}, 
\begin{equation}
J_k=R^nJ\, ,\quad k=2^n(2m+1)\, ,\quad n,m=0,1,\cdots\; ,
\label{e10}
\end{equation}
with $0<R<1$. The eigenvalue problem for $T$ corresponds to the second order
difference equations 
\begin{equation}
T_{j,j-1}V(j-1)-\Lambda V(j)+T_{j,j+1}V(j+1)=0\; ,
\label{e11}
\end{equation}
where $j=1,\cdots,2L$. To construct an exact recursion we eliminate from these
equations components of the form $V(4l+2)$, $V(4l+3)$ which are connected to a
$J$ coupling (indicated by crosses in Fig.~1a). After such a decimation the
triplet $(h,J,h)$ is replaced by a renormalized field $h'$ and keeping $R$
unchanged, the remaining couplings become $J'=RJ$ due to the hierarchical
structure of the sequence. Thus the renormalized equations keep the original
form with $\Lambda$ changed into $\Lambda'$. Introducing the reduced variables
$\lambda=J/h$ and $\widehat{\Lambda}=\Lambda/h$ one arrives at the
two-parameters recursion:   \begin{equation} 
\widehat{\Lambda}'={\widehat{\Lambda}\over\lambda}\,
(\widehat{\Lambda}^2-\lambda^2-1)\;
,\qquad\lambda'=R\,(\widehat{\Lambda}^2-\lambda^2)\; . 
\label{e12} 
\end{equation}
The RG-transformation has two non-trivial fixed points, governing the scaling
of the eigenstates at the top of the spectrum (DW) and at $\Lambda=0$ (IM),
respectively. The line $\widehat{\Lambda}=0$, corresponding to the IM
situation, is invariant under the RG transformation in Eq.~(\ref{e12}). Along
this line, starting with a ferromagnetic model with $\lambda>0$, after one
recursion step the system is transformed into an antiferromagnetic model with
$\lambda<0$. The critical IM with $\lambda_c=1/R$ is transformed into the IM
fixed point, which is situated at $\lambda^*=-1/R$. At the IM fixed-point the
leading eigenvalue of the transformation is $\epsilon_1=R+1/R$ and the
anisotropy exponent of the hierarchical IM is given by:
\begin{equation}
z=y_{\Lambda}={\ln(R+1/R) \over \ln 2}\; .
\label{e13}
\end{equation}
Thus scaling in the hierarchical IM close to the critical point
is essentially ani\-so\-trop\-ic.

Scaling of the eigenstates at the top of the spectrum is governed by the
DW fixed-point situated at
\begin{equation}
\lambda^*={R \over 1-R}\; ,\qquad\widehat{\Lambda}\,^*={\sqrt{1-R+R^2}\over
1-R}\; , 
\label{e14}
\end{equation}
and the leading eigenvalue is given by:  
\begin{equation}
\epsilon_1={1\over R}+R+{1\over 2}+\left[\left({1\over R}+R+{1 \over 2}
\right)^2-2\right]^{1/2}\; . 
\label{e15}
\end{equation}
Thus the wandering exponent of the walk is:
\begin{equation}
w={1\over y_{\Lambda}}={\ln 2\over\ln\epsilon_1}\; .
\label{e16}
\end{equation}
In the homogeneous model, with $R=1$, the DW fixed point is shifted to 
infinity since $\widehat{\Lambda}^*\to\infty$, $\lambda^*\to\infty$ and along
the separatrix $\widehat{\Lambda}/\lambda\to1$. To evaluate the scaling
behavior we introduce new variables: $\kappa=1/\lambda$ and
$\Delta=(\widehat{\Lambda}/\lambda)^2-1$, in terms of which the fixed point is
given by $\kappa^*=0$ and $\Delta^*=0$. Then the separatrix is a straight line:
$\Delta(\kappa)=a^* \kappa$, with $a^*=2$, and according to Eq.~(\ref{e12}) one
point of the $(\kappa,\Delta)$ plane with $\Delta=a\kappa$ will transform into
$(\kappa'=\kappa/a,\Delta'=(1-2/a^2)\Delta)$. Thus $a'=a^2-2$ and the leading
eigenvalue of the transformation is $\epsilon_1=4$, consequently
$w=1/y_{\Lambda}=1/2$, in agreement with known results~\cite{fisher86}. We note
that $w(R)$ is discontinuous at $R=1$, since from Eq.~(\ref{e16})
$\lim_{R\to1_-}w(R)<1/2$. 

\paragraph*{Period-doubling sequence}
In our next example, the couplings $J_k$ are generated according to the
period-doupling sequence~\cite{collet80} which follows from the substitution
$A \to AB$ and $B \to AA$. Here and in the following, the couplings
are parametrized as $J_A=J$ and $J_B=RJ$.

In an exact RG transformation, six sites out of eight have to be decimated, as
indicated on Fig.~1b. Associating new couplings $h'$ with the decimated blocks
one obtains a recursion in terms of $h'$ and $\Lambda'$  while
$\lambda$ and $R\lambda$, thus the ratio $R$, remain unchanged.
In terms of the reduced parameters the RG-transformation reads as
\begin{equation}
\widehat{\Lambda}'={\widehat{\Lambda}\over R\lambda^3}(c-d)\; ,\qquad\lambda'
={c\over R\lambda^2}\; ,
\label{e17}
\end{equation}
with $c=\widehat{\Lambda}^2(-\widehat{\Lambda}^2+1+\lambda^2)^2-R^2
\lambda^2(\widehat{\Lambda}^2-\lambda^2)^2$ and
$d=(\widehat{\Lambda}^2-1)^2-\lambda^2\widehat{\Lambda}^2(1+R^2)+
\lambda^2(1+R^2\lambda^2)$. The IM-fixed point of the transformation is at
$\widehat{\Lambda}^*=0$ and  $\lambda^*=-R^{-1/3}$, with the leading eigenvalue
$\epsilon_1=(R^{1/3}+R^{-1/3})^2$. Since the rescaling factor of the 
transformation is $b=4$ we obtain  
\begin{equation}
z={\ln(R^{1/3}+R^{-1/3})\over\ln 2}
\label{e18}
\end{equation}
for the anisotropy exponent of the period-doubling IM.

The top of the spectrum, corresponding to the DW problem, scales to a fixed
point with $\widehat{\Lambda}\to\infty$, $\lambda\to\infty$,  but
$\widehat{\Lambda}/\lambda\to R$. In terms of the variables $\kappa=1/\lambda$ 
and $\Delta=(\widehat{\Lambda}/\lambda)^2-R^2$ the fixed-point is at
$\kappa^*=0$ and $\Delta^*=0$, while the separatrix, close to the fixed point,
is of the form $\Delta(\kappa)=a^*\kappa^2+O(\kappa^4)$, with
$a^*=(\sqrt{2}R-2R^2)/(1-R^2)$. Then, according to Eq.~(\ref{e17}), a point of
the ($\kappa,\Delta$)-plane with $\Delta=a\kappa^2$ will transform to
$(\kappa'\sim\kappa^2,\Delta'\sim\Delta^{1/2})$. This type of scaling 
behavior is compatible with an essential singularity in the gaps at the top of
the spectrum,  
\begin{equation}
\Delta\Lambda_i\sim\exp(-CL^{\sigma})
\label{e19}
\end{equation}
with $\sigma=1/2$, since the rescaling factor is $b=4$. Thus the parallel
correlation length of the DW is given by $\xi_{\parallel}^{\rm DW}\sim\exp (C
L^{1/2})$ and the transverse fluctuations of the walk grow anomalously, on a
logarithmic scale:
\begin{equation}
\left< \left[ X(t)-X(0) \right]^2 \right> ^{1/2} \sim \ln^2(t)\; .
\label{e20}
\end{equation}
Here $X(t)$ denotes the position of the walker at time $t$. We note that the 
same asymptotic behavior is found in the Sinai model~\cite{sinai82} of a
one-dimensional random walk in a random environment. 

\paragraph*{Three-folding sequence} 
The three-folding se\-quence is generated by the substitutions
$A\to ABA$, $B\to ABB$ \cite{dekking83}. In the RG transformation - as
indicated on Fig.~1c - blocks of four sites are decimated out. Due to the
asymmetric nature of the blocks, after one RG step the transfer matrix becomes
asymmetric, too: $T_{j,j+1}/T_{j+1,j}=s$ for $j$ even, while
$T_{j,j+1}/T_{j+1,j}=s^{-1}$ for $j$ odd.

The recursion relations in this case are more conveniently expressed using
the variables $\widetilde{\Lambda}=\Lambda/J$, $\mu=h/J$ and
$s$, while $R$ remains unchanged:
\begin{equation}
\widetilde{\Lambda}'=\widetilde{\Lambda}
\left[\left(1-{c\over e}\right)
\left(1-{d\over e}\right)\right]^{1/2}\; ,\quad
\mu'=\mu^3{R\over e}\; ,
\label{e21}
\end{equation}
with $c=\mu^2(\widetilde{\Lambda}^2-\mu^2-R^2)$, 
$d=\mu^2(\widetilde{\Lambda}^2-\mu^2-1)$ and
$e=(\widetilde{\Lambda}^2-1)(\widetilde{\Lambda}^2
-R^2)-\mu^2\widetilde{\Lambda}^2$. We
note that the asymmetry parameter $s$, such that $s'=s(c-e)(d-e)$, 
does not enter into the recursions for $\widetilde{\Lambda}$ and $\mu$.

At the IM fixed point ($\widetilde{\Lambda}^*=0,~\mu^*=R^{1/2}$) 
the leading eigenvalue of the RG transformation is
$\epsilon_1=[(2+R)(2+R^{-1})]^{1/2}$, thus the anisotropy exponent is given by
\begin{equation}
z={\ln(2+R)(2+R^{-1})\over 2\ln 3}\; . 
\label{e22}
\end{equation}

The DW fixed point is again at infinity: $\widetilde{\Lambda}^*
=\infty$, $\mu^*=\infty$, with $\widetilde{\Lambda}^*/\mu^*=1$. 
The scaling behavior at this fixed point is similar to that in the 
period-doubling case. The eigenvalues at the top of the spectrum show an
essential singularity like in Eq.~(\ref{e19}) with $\sigma=1/2$ and the
transverse fluctuations grow on a logarithmic scale as in Eq.~(\ref{e20}).

\paragraph*{Paper-folding sequence}
Finally, we consider the pa\-per-fol\-ding se\-quence \cite{dekking83} which is
generated by the two-letter substitutions $AA\to AABA$, $AB\to AABB$,
$BA\to ABBA$ and $BB\to ABBB$. In the RG transformation,
decimating out blocks of two sites (Fig.~1d), alternating field variables $h_1$
and $h_2$ are generated for odd and even lattice sites, respectively. 
Furthermore, the transfer-matrix becomes asymmetric and the asymmetry 
parameters are different for odd and even elements.
As a consequence, the exact RG transformation contains altogether
six parameters. Here we just present the scaling behavior at the two
non-trivial fixed points, details of the calculation will be presented
elsewhere~\cite{igloi-turban}.

At the IM fixed point, the anisotropy exponent is continuously varying and
given by:
\begin{equation}
z={\ln(1+R)(1+R^{-1})\over\ln 4}\; .
\label{e23}
\end{equation}
At the DW fixed point the scaling is again of the streched exponential  form
with a leading behaviour for transverse fluctuations given by Eq.~(\ref{e20}).

Let us now turn to a discussion of the critical behavior we have obtained for
the IM and the DW. All the aperiodic IMs we considered are strongly anisotropic
with a continuously varying anisotropy exponent. In the extended parameter
space there is a line of fixed points parametrized by the coupling ratio $R$.
The critical behavior of the DWs is also found to be anomalous: the lines of
fixed points of the non-periodic systems are disconnected from the fixed point
of the homogeneous ones. For the hierarchical model the wandering exponent is
discontinuous at $R=1$ whereas for the other sequences the transverse
fluctuations grow on the same logarithmic scale.

The difference between the IM and the DW on the sa\-me lat\-ti\-ce can be 
understood using a relevance-irrelevance criterion~\cite{luck93} which is a
counterpart for aperiodic systems of the Harris criterion~\cite{harris74} for
random ones. The cross-over exponent associated with a layered non-periodic
perturbation is~\cite{luck93} $\phi=1+\nu(\Omega-1)$
where $\nu$ is the exponent of the correlation length, {\it perpendicular to 
the layers}, for the unperturbed system and $\Omega$ is a wandering
exponent~\cite{queffelec87} which characterizes the fluctuations in the 
couplings $J_k$ around their average $\overline{J}$ as
\mbox{$\sum_{k=1}^L(J_k-\overline{J})\sim L^{\Omega}$}. 

All the non-periodic sequences we considered have
$\Omega=0$.  For the IM with $\nu=1$ the cross-over exponent vanishes. Thus the
perturbation is marginal, which explains the continuous variation of the 
anisotropy exponent. On the other hand, for the anisotropic DW with
$\nu_\perp=1/2$, the cross-over exponent is $\phi=1/2$ and the perturbation is
relevant. This is again in agreement with our results.

\acknowledgements This work has been supported by the C.N.R.S.
and the Hungarian Academy of Sciences through an exchange program and by the
Hungarian National Research Fund under grants No OTKA TO12830 and No OTKA 
TO17485. FI is indebted to D. Sz\'asz for useful discussions on
Ref.~\cite{sinai82}. The Laboratoire de Physique du Solide is Unit\'e de
Recherche Associ\'ee au C.N.R.S. No 155.

\end{document}